\documentclass[conference]{IEEEtran}

\usepackage{mathtools}
\usepackage{tikz}
\usetikzlibrary{shapes,positioning,arrows}
\tikzstyle{block} = [rectangle, draw, fill=blue!20, text width=7em, text centered, rounded corners]
\usepackage{algorithm}
\usepackage{algorithmic}

\usepackage{placeins}

\title{AUV Optimal Path for Leak Detection}
\author{\IEEEauthorblockN{Olivier~Marceau}\IEEEauthorblockA{DCNS Research\\5 rue de l'Halbrane\\ F-44340 Bouguenais\\}\and\IEEEauthorblockN{Jean-Michel~Vanpeperstraete}\IEEEauthorblockA{DCNS Research\\280 avenue Aristide Briand\\F-92220 Bagneux}}

\begin{document}
	\maketitle
	
	\begin{abstract}
		This paper studies an optimal autonomous underwater vehicule (AUV) path planning method for both reducing average delay before pollutants detection in underwater mining, oil or gas fields and reducing AUV occupancy time. 
		The proposed technique, based on the \emph{bayesian search theory} framework and \emph{multi-objective optimization}, extracts optimal boustrophedon paths for leak detection in complex environment. We describe a multi-objective nonlinear mixed integer optimization model for both reducing global nondetection probability and path duration. We then propose a hierarchical algorithm combining two functions. The main function is a multi-objective cross entropy which places the tracklines. The second function sets the optimal speeds on each trackline by means of an interior point method. Numerical simulations show that the proposed framework is a very promising approach because the optimal paths cross spill of highly probable leaks before less probable ones. We show that our optimized paths outperform boustrophedon paths of same duration with uniform speed and spacing of trackline. Thanks to Pareto efficiency approach, our tool propose optimal trajectories for numerous AUV autonomies. Hence it can be used for both real time path planning and design purpose.

	\end{abstract}
	
	\section{Introduction}
	
	Inspection of underwater mining fields for turbidity, oil or gas leaks is an expensive task both for manning and equipment costs. One way to reduce costs is the deployment of autonomous underwater vehicles (AUVs). However AUVs have limited power and are often assigned to numerous critical tasks. 
	
	In order to increase AUVs effectiveness, one must take into account every piece of a priori knowledge on the monitored field. For our purpose, such an initial knowledge is summarized by three information. First, positions of probable leak sources are induced by the topology of submarine exploitation: a leak usually occurs on pipes junction, digging points or high pressure tank. Second information is the local submarine conditions like streams, temperatures, pressures or densities which influence pollutants propagation. The last knowledge is the history of previous leak events and previous inspections.
	
	All these information can be combined to produce a simulation of pollutant propagation. See \cite{hyun2010coastal} and \cite{north2015influence} for oil propagation models. Such a simulation permits to define potential pollutant fill area for each identified leak source. Moreover expertise, previous events and previous inspections analysis allow to infer a priori probability occurrence of each leak. Fill areas and corresponding leak probabilities are designed by \emph{a priori leak map} throughout this article.
	
	Bayesian search theory permits to exploit a priori leak map for building a metric which evaluates quality of a trajectory. Such a theory aims to improve effectiveness of searching efforts in a constrained and uncertain environment. This discipline was introduced by the Antisubmarine Warfare Operations Research Group (ASWORG) during World War II \cite{koopman1946search}. Since that time, search theory became a widely used discipline in operations research; interested readers may consult extensive surveys \cite{benkoski1991survey} and specialized books \cite{stone1976theory,washburn2002search}. More recently, search theory was apply on the search of flight AF447 Rio - Paris \cite{stone2011search}. 
	According to \cite{le2000searching}, a classical search theory problem is characterized by three data sets:
			\begin{itemize}
				\item the probabilities map of the searched object in various possible location,
				\item the local detection probability that an amount of local search effort could detect the target,
				\item the total amount of searching effort available. 
			\end{itemize}
	In case of pollutant leak detection, the first set is the a priori leak map, and third set is the total amount of time available on the AUV.  
	Second data set is generally more complex because it requires a precise sensor measurement characterization. For our purpose we choose a rather simple sensor model consisting in a continuous local point measurement of characteristic time $\tau$ such that probability of detection of a pollutant if the sensor stay a duration $\Delta t$ in the leak fill area is given by $1-exp(-\frac{\Delta t}{\tau})$.
	\begin{figure}[ht]
		\centering
		\includegraphics[width=0.9\columnwidth]{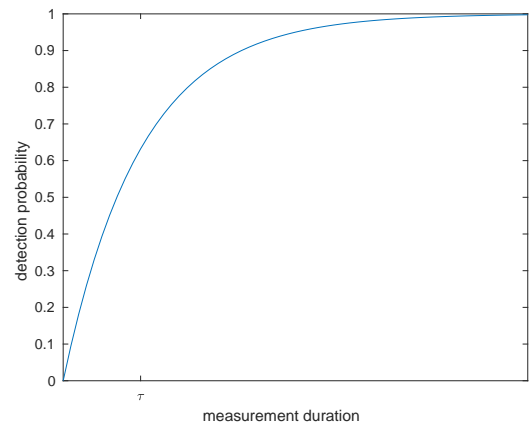}
		\caption{Pollutant detection probability in function of time spent in the spill area.}
	\end{figure}
		
   Our aim is also to reduce time spent by the AUV on the area to let this AUV fulfill other tasks. For example one can say staying one more hour in the mining field is worthless if global leak detection probability only increases of 0.1\%.
		The conventional method of handling multiple objectives is to construct a combined objective and thus solve a single objective problem. See \cite{negre2012stochastic,yang2004decentralized} for example of such methods. However, there are several drawbacks to this. First, since different objectives have different meanings, the combined objective is difficult to interpret and validate. Second, in combining several objectives in a single function, one must know the relative importance of each objective.
		
		In this paper, a multiobjective optimization (MOO) framework exhibits a set of solutions rather than a single solution. Each solution makes a compromise among multiple and often conflicting objectives. Such a set of solutions is commonly known as a Pareto-optimal set in which Pareto-optimality is defined in terms of a dominance relation between two solutions as follows: given two solutions $u$ and $v$, $u\neq v$, $u$ is said to dominate $v$ if $u$ is not worse than $v$ in all objectives and $u$ is strictly better than $v$ for at least one objective. For example, for a minimization problem
		\begin{equation}
			\min_{x \in \mathcal{X}} \mathbf{f}(x)=\left(\mathrm{f}_1(x),\mathrm{f}_2(x),\dots,\mathrm{f}_K(x)\right)
		\end{equation}
		solution $u$ is better than $v$ with respect to objective $i$ if $\mathrm{f}_i(u) \leq \mathrm{f}_i(v)$ and $u$ is said to dominate $v$ denoted as $u \prec v$. One may consult \cite{ehrgott2006multicriteria} for increasing knowledge on MOO.
		Hence Pareto set is the set of non dominated solutions.
		
	AUV path planning is a widely studied subject: \cite{petres2005underwater} describes a method to find minimal path duration to join an arrival point in presence of obstacles and streams; \cite{soulignac2006fast} proposes a traveling salesman problem (TSP) approach to minimize flight time path for multiple site surveillance.
	Multiobjective path planning with exploration of the Pareto frontier has previously been studied in \cite{tian2008multi} in a context of target surveillance by AUVs. 
	
	\section{Problem formulation}
	
\begin{figure}[ht]
		\centering
		\includegraphics[width=0.9\columnwidth]{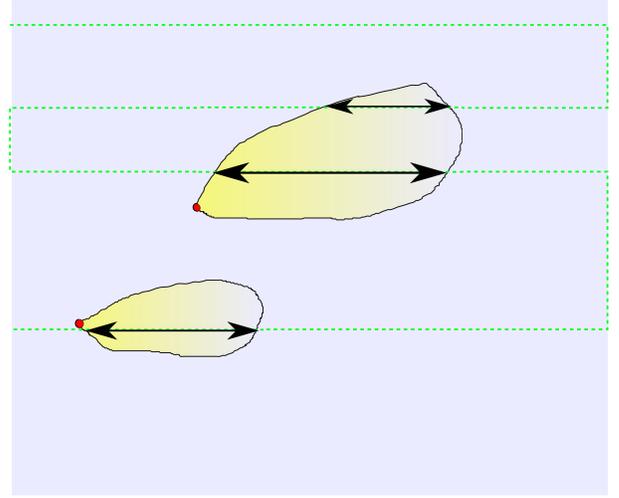}
		\caption{Boustrophedon pattern in green dashed line among two spill of pollutant emanating from two sources (red points).}
		\label{fig:boustro2}
	\end{figure}

	Before any mathematical modeling, we make several preliminary assumptions. First, all trajectories are calculated at a constant depth. One can adapt our method by applying it to several possible depths and then merging each Pareto front and pruning dominated trajectories. Thus, without loss of generality, our path planning problem is reduced to a planar one. Then we restrict possible patterns to classical boustrophedon ones (figure \ref{fig:boustro2}) and we enforce constant speed on each trackline so the decision variables are the ''horizontal'' trackline positions and the AUV speed on the corresponding trackline and we neglect streams.
	
	Let $i=1,\dots,n$ denote potential leak sources. Each source $i$ is associated with an a priori probability of leak $\pi_i$ and with a spill area $A_i$. In order to maximize leak detection, our AUV must spend maximal time in each leak spill area. We restrain possible tracklines to a finite set of $m$ well chosen lines. Hence boustrophedon pattern consist in a selection of "horizontal" tracklines in a set denoted by $j=1,\dots,m$ and we define $v_j$  the AUV speed on trackline $j$ and $\delta_j$ the integer variables representing the number of traveling on the trackline $j$. The $\delta_j$ take their values between $0$ and $z$, with $z$ the maximal number of times the AUV can travel a single trackline. With such notations the classical search theory non detection probability can be formulated
	\begin{equation}\label{eq:crit_pnd}
		P_{ND}=\sum_{i=1}^n \pi_i \cdot \exp{\left(-\frac1\tau \sum_{j=1}^m \frac{l_{i,j}}{v_j}\delta_j\right)}
	\end{equation}
	with $l_{i,j}$ the length of the trackline $j$ inside the spill area $A_i$ and $\tau$ the characteristic time of the mobile sensor. Path duration can be expressed
	\begin{equation}\label{eq:crit_time}
		T=\sum_{j=1}^m \frac{l_j}{v_j}\delta_j
	\end{equation}
	with $l_j$ the length of trackline $j$. One can notice that \eqref{eq:crit_pnd} and \eqref{eq:crit_time} neglect "vertical" pieces of the pattern.
	
	AUV physical capacities and sensor requirements impose $v_{\min} \leq v_j \leq v_{\max}$ for all $j=1,\dots,m$ and 
	\begin{equation}\label{eq:const_time}
		\sum_{j=1}^m \frac{l_j}{v_j}\delta_j \leq T_{\max} \; .
	\end{equation}
	
	Finally our problem can be summarized
	\begin{equation}\label{eq:prob}
		\left[ \begin{array}{l}
		\begin{array}{l}
		\min_{\boldsymbol{\delta},\mathbf{v}}	\sum_{i=1}^n \pi_i \cdot \exp{\left(-\frac1\tau \sum_{j=1}^m \frac{l_{i,j}}{v_j}\delta_j\right)}\\
		\min_{\boldsymbol{\delta},\mathbf{v}} \sum_{j=1}^m \frac{l_j}{v_j}\delta_j
		\end{array}\\
		s. t. \left\{\begin{array}{l}
		\forall j=1,\dots,m, \; \delta_j \in \{0,z\}\\
		\forall j=1,\dots,m, \; v_{\min} \leq v_j \leq v_{\max}\\
		 \sum_{j=1}^m \frac{l_j}{v_j}\delta_j \leq T_{\max}
		\end{array}
		\right.
		\end{array}\right].
	\end{equation}
	\eqref{eq:prob} is a Mixed Integer Programming (MIP) nonlinear biobjective problem.
	
	Search theory also gives a mean to update distribution of leak probability considering no detection occurs: immediately after the AUV moves, updated leak probabilities $\pi'_i$ are given by formula
	\begin{equation}\label{eq:a_posteriori}
	\pi'_i = \pi_i \exp{\left(-\frac1\tau \sum_{j=1}^m \frac{l_{i,j}}{v_j}\delta_j\right)}
	\end{equation}

	\section{Algorithm description}

		We propose a hierarchical approach to solve problem \eqref{eq:prob}: a main function select best tracklines and average speed while a secondary function optimize speed on each trackline.
		
		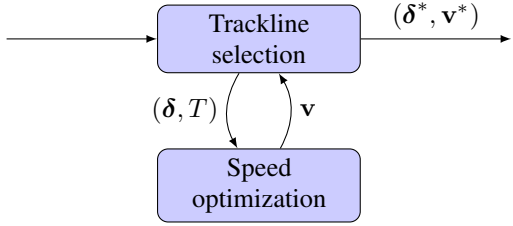
\begin{figure}[ht]
			\centering
			\begin{tikzpicture}[node distance = 1cm and 2cm,>=latex]
			\node(entree){};
			\node(master) [block,right=of entree]{Trackline selection};
			\node(slave) [block,below=of master]{Speed optimization};
			\node(sortie) [right=of master]{};
			\draw[->] (entree) -- (master) node[near start,above]{};
			\draw[->] (master) to[bend right]node[midway,left]{$(\boldsymbol{\delta},T)$} (slave);
			\draw[->] (slave) to[bend right]node[midway,right]{$\mathbf{v}$} (master);
			\draw[->] (master) |- (sortie) node[near end,above]{$(\boldsymbol{\delta}^\ast,\mathbf{v}^\ast)$};
			\end{tikzpicture}
			\caption{Hierarchical architecture for AUV path planning.}
		\end{figure}
	
	This decomposition aims to divide one difficult problem into several easier problems while producing optimal solutions for problem \eqref{eq:prob}.
	
		\subsection{Speed optimization}

		Considering only the trackline speeds and assigning a global path duration $T$, problem \eqref{eq:prob} become
		\begin{equation}\label{eq:prob_speed} 
		P_{ND}^\ast(\boldsymbol{\delta},T)=\left[ \begin{array}{l}\scriptstyle
		\min_{\mu_j|\delta_j=1}	\sum_{i=1}^n \pi_i \cdot \exp{\left(-\frac1\tau \sum_{j=1}^m l_{i,j}\mu_j\right)}\\
		s. t. \left\{\begin{array}{l}\scriptstyle
		\forall j=1,\dots,m, \; \frac1{v_{\max}} \leq \mu_j \leq \frac1{v_{\min}}\\ \scriptstyle
		\sum_{j=1}^m l_j \mu_j  = T
		\end{array}
		\right.
		\end{array}\right].
		\end{equation}
		with $\mu_j$ the inverse of trackline speeds ($\mu_j:=\frac1{v_j}$ for all $j=1,\dots,m$). 
		
		Thanks to the substitution of $v_j$ by $\mu_j$, \eqref{eq:prob_speed} is a well-defined convex problem which can be easily solved by mean of an interior point method. Readers who want to improve knowledge in convex optimization and interior point method may consult \cite{byrd2000trust}.
	
		\subsection{Trackline selection optimization}
		
		\eqref{eq:prob_speed} and substituting discrete variables $\delta_{j}$ by binary variables $\delta_{j,k}$ for $k=1,\dots,z$ ($\delta_{j} = \sum_{k=1}^z  \delta_{j,k}$) permits to rewrite problem \eqref{eq:prob}:
		\begin{equation}\label{eq:master_prob}
		\left[ \begin{array}{l}
		\begin{array}{l}
		\min_{\boldsymbol{\delta},T} P_{ND}^\ast(\boldsymbol{\delta},T)\\
		\min_{\boldsymbol{\delta},T} T
		\end{array}\\
		s. t. \left\{\begin{array}{l}
		\forall j=1,\dots,m, \; \forall k=1,\dots,z, \; \delta_{j,k} \in \{0,1\}\\
		\sum_{j=1}^m \sum_{j=1}^m \frac{l_j}{v_{\max}}\delta_{j,k} \leq T \leq T_{\max}
		\end{array}
		\right.
		\end{array}\right].
		\end{equation}
		\eqref{eq:master_prob} is a biobjective mixed programing problem with with one nonlinear objective, $z \cdot m$ binary variables and one continuous variable $T$. Thus our problem is well-posed for applying the multiobjective cross-entropy method (MOCE) introduced by \cite{unveren2007multi} as an extension of the original cross-entropy method (CE) \cite{rubinstein2013cross}. The cross-entropy method is a stochastic learning algorithm inspired from rare event simulations. 
		
%
%

	\section{Results}
	
\begin{figure}[ht]
		\centering
		\includegraphics[width=0.9\columnwidth]{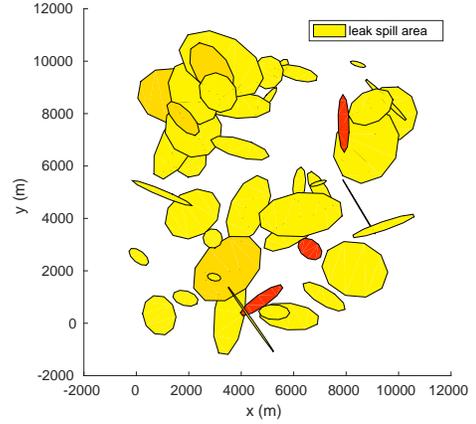}
		\caption{Target area with 50 potential leak sources. Yellow leak surfaces are less probable ones and red are the most probable.}
		\label{fig:area}
	\end{figure}

	Our algorithm was implemented in MATLAB and tested on simulated data on a square area of size 10 km x 10 km including 50 potential leak sources, 42 with a priori probability of 0.05, 5 with an a priori of 0.15 and 3 with an a priori of 0.80 (see figure \ref{fig:area}). Each leak area is the convex hull of an ellipse with random parameters. One potential trackline is created on each extrema of every leak area, which conducts to 100 potential tracklines. The AUV has an autonomy of 10 hours and can move between 2 and 5 knots. The sensor has a characteristic time of 200 seconds. Given that the AUV can range less than 20 tracklines, the number of possible boustrophedon paths which must be explicitly enumerated is approximatively $10^{40}$.
	
\begin{figure}[ht]
		\centering
		\includegraphics[width=0.9\columnwidth]{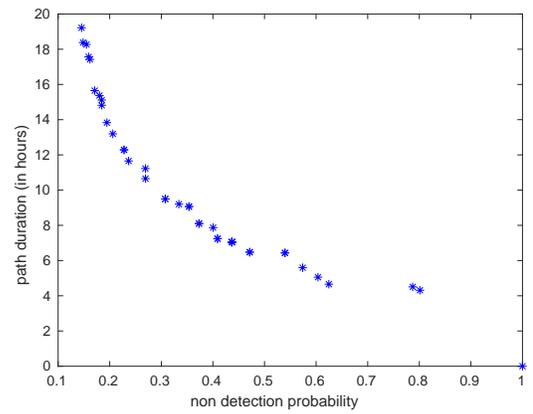}
		\caption{Pareto set.}
		\label{fig:pareto}
	\end{figure}

	By applying the proposed algorithm on such a test case, we obtain a pareto set of thirty solutions given on figure \ref{fig:pareto}.
	Our methods takes approximatively fifteen minutes on a standard desktop computer.

\begin{figure}[ht]
		\centering
		\includegraphics[width=0.9\columnwidth]{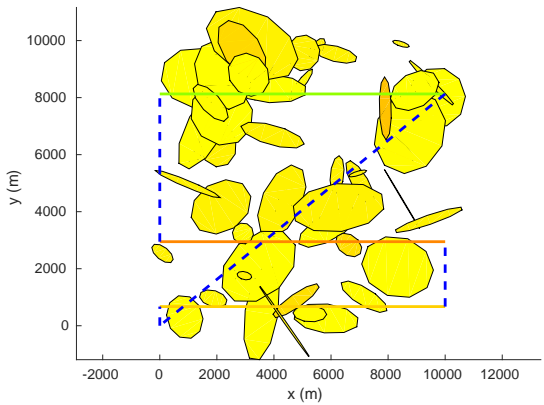}
		\caption{Example of a trajectory with updated leak probabilities, non detection probability of 0.4 and path duration of 7.5 hours. Green tracklines are traveled at lowest speed while red trakclines are traveled at highest speed. Blue dashed tracklines are neglected for detection.}
		\label{fig:traj}
	\end{figure}

	For a given trajectory, one can estimate updated leak probabilities according to \eqref{eq:a_posteriori}. Figure \ref{fig:traj} shows such a trajectory with updated leak probabilities.
	This example shows that our algorithm choose trajectories which cross first the fill area of high leak probability.
	
	On a design point of view, the bicriteria Pareto Front can be modified in a detection probability curve in function of path duration. Moreover one can easily compare 1 AUV versus 2 AUVs configuration. We also add to this comparison result of a non optimized path consisting of boustrophedon with regularly spaced tracklines at a constant speed on full path.
	\begin{figure}[ht]
	\centering
	\includegraphics[width=0.9\columnwidth]{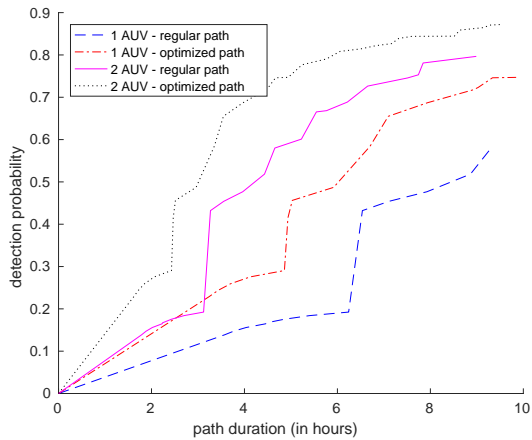}
	\caption{Comparison between regularly spaced and optimized path with one or two AUVs.}
	\label{fig:comp}
	\end{figure}
	Figure \ref{fig:comp} illustrates this approach and show that optimized 1 AUV trajectories almost reach regular 2 AUVs trajectories. The noticeable break on each curve matches with transition between one low speed trackline patterns and two high speed tracklines patterns. Transitions between higher number of tracklines are also apparent but less significant. Note that our approach for $n$ AUVs only consists in increasing path duration by a factor $n$. The optimal tracklines - AUVs assignment is a difficult problem  which is beyond this article scope. 

	\FloatBarrier
	\section{Conclusion}
	
	This article proposes a new method for AUV path planning for early leak detection in mining or oiling fields. Numerical experiments show that optimized paths outperform classical regularly spaced boustrophedon paths. Thanks to a bicriteria and time effective approach, our algorithm can serve purpose of both real time decision aid and design of a new system. Furthermore our optimized path can reduce number of AUVs required to monitor an area and thus can highly reduce initial and maintenance costs of a monitoring system.
	 
	Future works may develop several aspects. First our algorithm will be evaluated on realistic leak fill areas using existing propagation models \cite{hyun2010coastal} and \cite{north2015influence}. Then optimized paths could take into account submarine streams. Third, each kind of pollutant has is own significance, thus our approach should include one objective for each potential pollutant in the area instead of mixing all pollutants in the same objective . Another promising field could be the extension of our model to multiple heterogeneous AUVs.

	\FloatBarrier
	\bibliographystyle{IEEEtran}
	\bibliography{IEEEabrv,SSW}
	
\end{document}